\PassOptionsToPackage{table}{xcolor}

\documentclass[journal]{vgtc}                     

\onlineid{1307}

\vgtccategory{Research}

\vgtcpapertype{application/design study}

\title{Visual Analytics for Understanding Draco's Knowledge Base}

\author{%
  \authororcid{Johanna Schmidt}{0000-0002-9638-6344},
  Bernhard Pointner, and 
  \authororcid{Silvia Miksch}{0000-0003-4427-5703}
}

\authorfooter{
  \item
  	Johanna Schmidt and Bernhard Pointner are with VRVis Zentrum für Virtual Reality und Visualisierung Forschungs-GmbH.
  	E-mail: johanna.schmidt@vrvis.at.
  \item Silvia Miksch is with the Centre for Visual Analytics Science and Technology (CVAST) at TU Wien.
  	E-mail: silvia.miksch@tuwien.ac.at.
}

\abstract{%
  Draco has been developed as an automated visualization recommendation system formalizing design knowledge as logical constraints in ASP (Answer-Set Programming). With an increasing set of constraints and incorporated design knowledge, even visualization experts lose overview in Draco and struggle to retrace the automated recommendation decisions made by the system. Our paper proposes an Visual Analytics (VA) approach to visualize and analyze Draco's constraints. Our VA approach is supposed to enable visualization experts to accomplish identified tasks regarding the knowledge base and support them in better understanding Draco. We extend the existing data extraction strategy of Draco with a data processing architecture capable of extracting features of interest from the knowledge base. A revised version of the ASP grammar provides the basis for this data processing strategy. The resulting incorporated and shared features of the constraints are then visualized using a hypergraph structure inside the radial-arranged constraints of the elaborated visualization. The hierarchical categories of the constraints are indicated by arcs surrounding the constraints. Our approach is supposed to enable visualization experts to interactively explore the design rules' violations based on highlighting respective constraints or recommendations. A qualitative and quantitative evaluation of the prototype confirms the prototype's effectiveness and value in acquiring insights into Draco's recommendation process and design constraints.
}


\teaser{
  \centering
  \includegraphics[width=\linewidth]{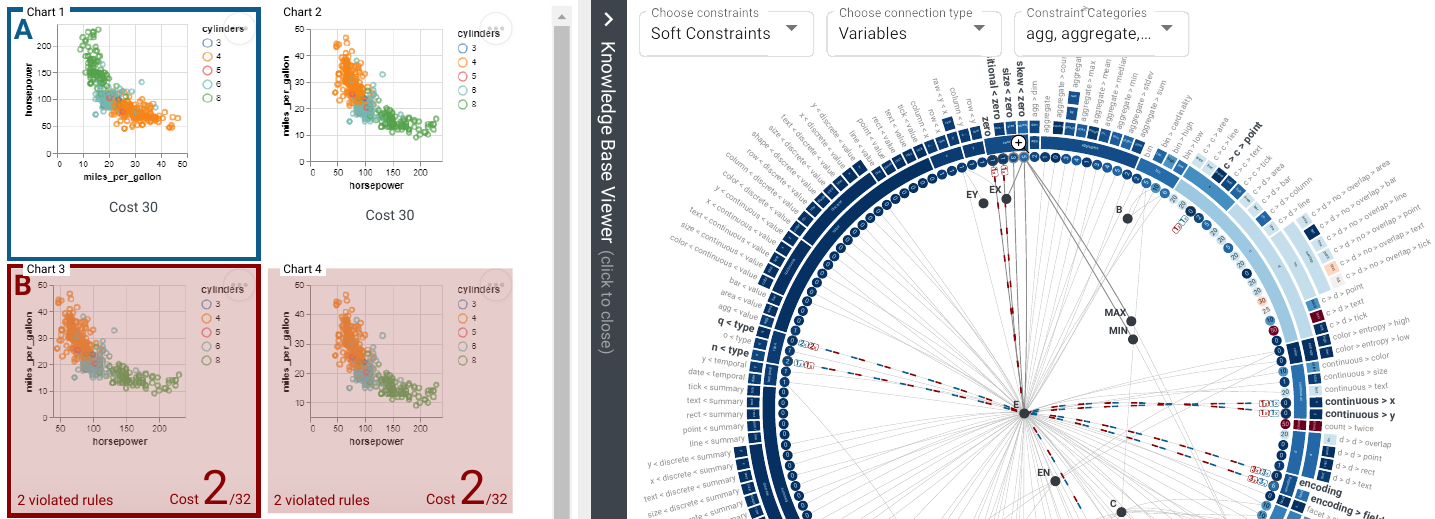}
  \caption{Visually inspecting Draco's recommendations. Our interactive Visual Analytics solution allows users to explore the set of rules the visualization recommendation system Draco is built on. On the left side, four recommended visualizations are shown. Every recommendation has costs assigned, which relates to how many rules have been violated by this recommendation. On the right side we present our hypergraph-based visualization of the set of rules and constraints that are used by Draco. By selecting recommendations on the left (A, blue, and B, red), the rules violated by these visualizations are highlighted in the graph (red and blue dashed lines).}
  \label{fig:teaser}
}

\graphicspath{{figs/}{figures/}{pictures/}{images/}{./}} 

\usepackage{tabu}                      
\usepackage{booktabs}                  

\usepackage{mathptmx}                  

\begin{document}

\firstsection{Introduction}
\label{sec:introduction}

\maketitle

\begin{figure}[t!]
	\centering
	\includegraphics[width=\columnwidth]{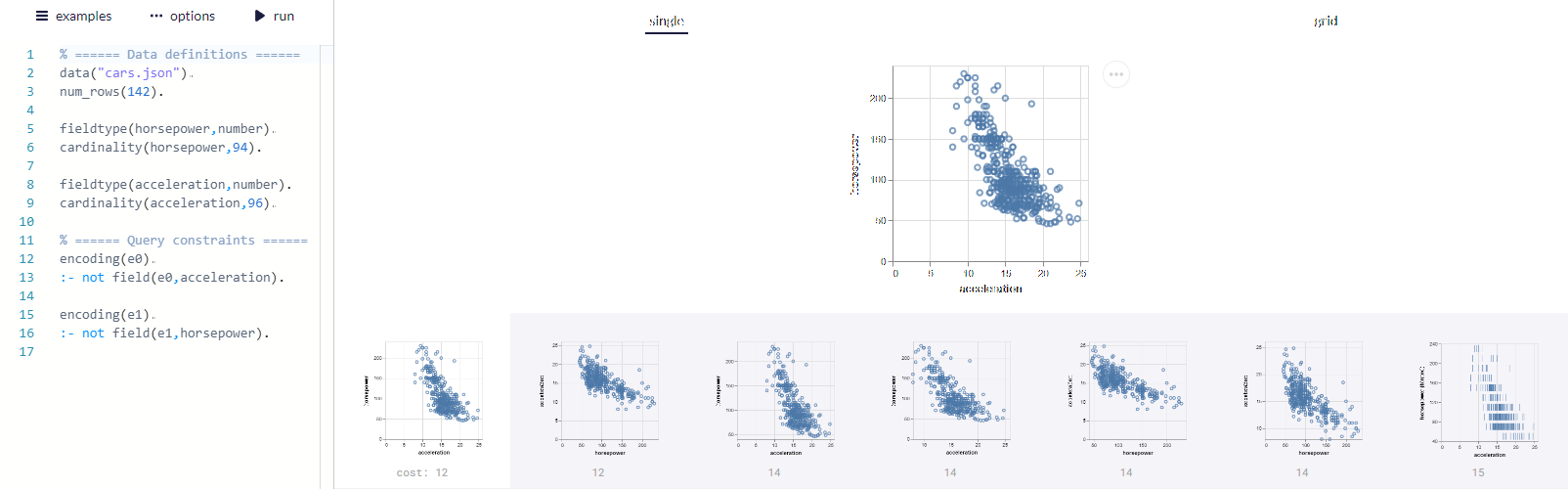}
	\caption{Draco editor. In the web-based Draco editor, users can specify data and attributes, and Draco will recommend fitting visualizations.}
	\label{fig:draco}
\end{figure}

Creating effective charts and plots is becoming a crucial task with today's increasing number of data and its complexity. Visual encodings, such as visual channels, mark types, binning and aggregation operations, faceted view, and others, must be carefully selected and used to visualize given data effectively. This knowledge of how to properly use and apply these encodings can be learned from books, scientific papers, courses in schools, and similar. Researchers argue that visualization research should go beyond manually curated and applied visualization design guidelines~\cite{corr/abs-1807-06641}. Design guidelines are continually evolving. Consequently, visualization recommendation systems for automated visualization design have enjoyed increasing interest. Visualization recommendation systems suggest proper visualizations (mostly charts/plots) for a given dataset, according to pre-defined rules. As one of the most prominent representatives, Moritz et al.~\cite{8440847} proposed \emph{Draco} in 2018. The Draco editor is shown in \autoref{fig:draco}. Based on the Vega-Lite grammar~\cite{7539624}, a high-level language for visualization, Draco defines a vast set of rules for recommending visualizations based on given data. Users can define datasets and constraints (e.g., the data domains they are interested in), and Draco will recommend fitting visualizations based on the rules that have been stored beforehand. In \autoref{fig:draco}, users provided contextual factors (such as column importance and task type). Draco uses predicates to condition the visualization of scatter plots with color to include a third attribute.

The motivation for a thesis~\cite{pointner2022} and this paper arose from a detailed analysis of Draco's application to real-world problems. We specifically concentrated on Draco since it is freely available (compared to Excels or Tableau's recommendation principles) and is one of the success stories in the visualization research community and, therefore, is well-known. As it can be seen in \autoref{fig:teaser} (left) and \autoref{fig:draco}, Draco informs the user about how much a recommended visualization violates the given constraints (i.e., \emph{costs} indicated below every recommendation). Costs can be used as a hint of how well the visualization fits the data. Apart from the costs, the rules that led to a visualization being considered not optimal are not revealed to the user. As such, Draco currently lacks a deeper understanding of how and which rules influence the recommendation process, and which factors lead to a visualization being considered optimal (or not). This leads to the fact that recommended visualizations cannot be directly compared, and specific soft constraints' influence on these recommendations is unclear. From a visualization designer's perspective, it is impossible to get an overview of which rules are already included in the systems and their interconnections. Draco uses Answer Set Programming (ASP) to interpret its rules (see \autoref{sec:terms} for a more detailed description). This means that when wanting to work with the set of rules and guidelines directly, visualization designers would have to make themselves familiar with logic programming in general and ASP in particular. The rules are only textually encoded and no advanced filter or reorganization mechanisms are available. Due to the lack of a visual interface, the current implementation of Draco requires the user to know how to formulate logical expressions syntactically correctly, which parameters are already encoded into the system, and how the expressions relate to each other. It is, therefore, a very time-consuming task to investigate the system's ingredients and purpose.

We believe that new Visual Analytics approaches toward understanding Draco's recommendation system's set of rules and recommendation process should be prioritized. It is helpful for Draco users and visualization designers to understand visualization recommendations better, and it will open up possibilities for the visual representation of rule-based systems in general. This also reflects the current discussions on explainable AI~\cite{BARREDOARRIETA202082} and, similarly, on explainable recommendation systems~\cite{10.1145/506443.506619}. Making Draco more accessible and understandable by researchers and visualization designers would create the basis for further developments. To address these gaps, we propose a \textbf{Visual Analytics (VA) approach to understanding rule-based recommendations} of the visualization recommendation system Draco. Our proposed approach aims to investigate the internal set of rules of Draco and find proper tooling and visualization to support developers, researchers, and designers in better understanding the system.

\begin{itemize}
	
    \item We propose \emph{adding meta-data to Draco's constraints and visualizing them} to reveal the underlying interrelationships using a hypergraph structure inside the radial-arranged layout. This helps visualization experts gain the desired insights into the underlying set of rules (\autoref{sec:approach}).
    
    \item We propose \emph{visually connecting the recommendation results with the constraints and highlighting respective violations}. This enables experts to adjust the constraints and their weights accordingly to concrete preferences (\autoref{sec:approach}).
    
    \item The \emph{effectiveness} of our proposed VA concepts is shown by applying it to use cases (\autoref{sec:approach}) and by evaluating it quantitatively and qualitatively using performance analyses, heuristic-based evaluation methodology, and interviews (\autoref{sec:evaluation}).
    
\end{itemize}

\section{Related Work}
\label{sec:relatedwork}

\paragraph{Visualization recommendation systems}

Several integrations and applications of data visualization recommendation systems have been proposed in the last years~\cite{10.1007/978-3-030-26636-3_1}, with Draco~\cite{8440847} as one of the most prominent examples. Recommendation systems have also already been incorporated into publicly available and known data analysis tools, like Tableau's \emph{Show Me} feature~\cite{4376133} and Google's \emph{Explore} tool~\cite{EP3491538A1}. Four types of recommendation systems can be distinguished~\cite{ivappkaur17}: Data characteristics-oriented, task-oriented, domain knowledge-oriented, and user preference-oriented. This categorization was extended by Vartak et al.~\cite{10.1145/3092931.3092937} with the field of `visual ease of understanding.' The majority of the most relevant visualization recommendation systems can be categorized as data characteristics oriented. Such systems recommend visualizations based on the characteristics of the data. In contrast, task-oriented approaches focus on the user's intent respectively goal. For instance, \emph{BOZ}\cite{10.1145/108360.108361} analyses tasks encoded as logical facts and produces graphics with perceptional instructions to reach the goal. \emph{IMPROVISE}~\cite{conf/cmc/ZhouF98} performs a natural language query on the user's intents to extract and automatically propose visual tasks. Other task-oriented systems are \emph{HARVEST}~\cite{10.1145/1502650.1502695} or the more recent \emph{DataSlicer}~\cite{corr/AlborziCCDHPRS17}.

We can further distinguish between approaches recommending \emph{what} data to show (e.g., automatic feature ranking~\cite{10.1057/palgrave.ivs.9500091} or view suggestions~\cite{7498285}) in contrast to \emph{how} to show data giving recommendations for charts and plots (e.g., \emph{Voyager 2}~\cite{10.1145/3025453.3025768}, \emph{DeepEye}~\cite{8509240}, \emph{VizML}~\cite{conf/chi/HuBLKH19}, \emph{Data2Vis}~\cite{8744242}, and, as already mentioned, \emph{Draco}~\cite{8440847}). The underlying recommendation strategies differ by rule-based, machine learning-based, and hybrid approaches. Earlier recommendation strategies mainly encoded decisions on visualization guidelines and principles as a set of rules and constraints. The predominant part of systems still rely on hand-crafted rules. Recent approaches, like \emph{VizML}~\cite{conf/chi/HuBLKH19} and \emph{Data2Vis}~\cite{8744242}, increasingly use machine learning to create recommendations. These systems learn the relationships between design guidelines and principles and weight their influence on respective visual encodings. Draco~\cite{8440847} employs a hybrid approach that combines hand-crafted visualization facts as well as  hard and soft constraints with learned weights from a Support Vector Machine (RankSVM) model trained on labeled visualization pairs.

\textbf{Our focus for showing how Visual Analytics can be used to understand a recommendation system's knowledge base is on Draco~\cite{8440847}.}

\paragraph{Visualization for logic programming}

Since we focus on Draco~\cite{8440847} in this work, we explore on existing visualization approaches for understanding logical rules. The visualizations in the context of logic programming are very diverse and address different goals and steps. These steps reach from interpreting the logic programs using a proper visualization of the program to the visualization of the solvers' results. We especially concentrate on visualizations showing the program respectively the query encoded as logical expressions.

Dependency graphs are often used to show the relations between the units of a logic program. These graphs can be visualized as free node-link layout graphs, where the node positions are only dependent on their linked neighbor nodes or in tree respectively flow structures showing the deductions of the predicates. An example of free layout representing the dependencies of the units of a logic program in Prolog-style has been proposed by Hopfner et al.~\cite{1199218}. Their approach suggests a directed graph representing predicates as circles. Glyphs are then connected by arrow links showing the directions of the deductions. Another way to visualize logic programs in Prolog is by mapping the program to top-down trees~\cite{10.1145/888251.888258}. The development environment \emph{ASPIDE}~\cite{10.1007/978-3-642-20895-9_37} for answer-set programming integrates a visual editor for creating and modifying logic programs. The editor resembles a UML diagram typically used to describe classes and their properties in programming languages.

\textbf{Our approach builds upon existing ideas and employs a network visualization-based approach to visualize the knowledge base.}

\paragraph{Graph-based visualizations of set relationships}

There are almost uncountable graph-based approaches and applications for showing relationships between a set of entities. To get an overview, Nobre et al.~\cite{2019_eurovis_mvnv} categorizes them into \emph{node-link} layouts, \emph{tabular} layouts, and \emph{implicit tree} layouts. Node-link layouts are the most common graphical representation for graphs and networks. Schulz and Schumann~\cite{1648257} further subdivided such layouts into free layouts, where the nodes' positions are not restricted, styled layouts, where the positions follow a predefined scheme, and fixed layouts, where the position is determined by nodes' attributes, such as latitude and longitude. Examples of free layouts are force-directed layouts where a node's position depends only on the links to its neighbors. Styled layouts typically make use of predefined schemes, like grids or axis-parallel and radial arrangements~\cite{2019_eurovis_mvnv}. Besides, on-node and on-edge encoding is often used to encode additional information into graph visualizations.

The hair-ball problem mentioned by Schulz and Hurter~\cite{SchuHurt13} is typical in node-link layouts. When the number of nodes and links between those nodes exceeds a critical sum, the visualization suffers from cluttering and overplotting. To address this problem, researchers developed various approaches to reduce the clutter by aggregating or filtering either the nodes, the edges, or both. For example, edge bundling~\cite{10.1111/cgf.13213} and edge splatting~\cite{BurchBW12} are very commonly used techniques. A different approach would be the use of hyperedges. The edges of a completely connected subset of nodes of a graph are removed, a new hypernode is added, and this introduced node is newly connected with the original nodes of the subset~\cite{6634104}.

\textbf{We used the approach of hypergraphs~\cite{6634104} to solve the problem of having too much clutter in the visual representation of the rules and constraints.}

\paragraph{Visualizations for hierarchical data}

Hierarchical data, in general, can be visualized in many different ways. Schulz~\cite{6056510} suggests three design axis to describe such tree visualizations: dimensionality (2d, 3d, hybrid), edge representation (explicit, i.e., links, or implicit, e.g., placement), and node alignment (radial, axis-parallel, or free). Dimensionality referes to displaying the data in 2 or more dimensions. Edge representation can be done explicitly, e.g., tidy trees~\cite{6295787} also know as \emph{Dendograms}. Typical examples~\cite{885091} for implicit edge representations of hierarchical structures are icicle plots, sunburst charts, sundown charts, or treemaps. This list can be extended by circle packing approaches~\cite{10.1145/1124772.1124851} which is a subcategory of treemaps, and hybrid approaches, like phylogenetic trees~\cite{Nielsen2016}.

Most relevant related to our approach are visualization showing hierarchical structures in graphs. Vehlow et al.~\cite{10.1111/cgf.12872} categorized the visualization of hierarchical graph structures the approaches \emph{visual node attributes} (flat group structures), \emph{juxtaposed} (side-by-side), \emph{superimposed} (nested), and \emph{embedded} (aggregated) visualizations. Visualizations employing a visual node attributes approach mostly follow the principle of displaying node-link diagrams with different layouts~\cite{PURCHASE2002501} or matrix-based solutions~\cite{4126232}. Juxtapositioned approaches display the group information next to the graph, either separate~\cite{10.1007/s12650-014-0271-9} or attached~\cite{10.1111/j.1467-8659.2008.01213.x}. Superimposed approaches place all information in the same space~\cite{10.1111/cgf.12512}. Embedded solutions either include group information in node-link representations~\cite{10.1111/cgf.12400}, or use hybrid solutions (e.g., based on matrix representations) for displaying group information~\cite{4376154}.

\textbf{We also deal with hierarchical structures in our data domain. In our approach we followed an embedded~\cite{10.1111/cgf.12872} visualization approach to include the hierarchical information in the graph representation.}

\begin{figure}[b!]
	\centering
	\includegraphics[width=.9\columnwidth]{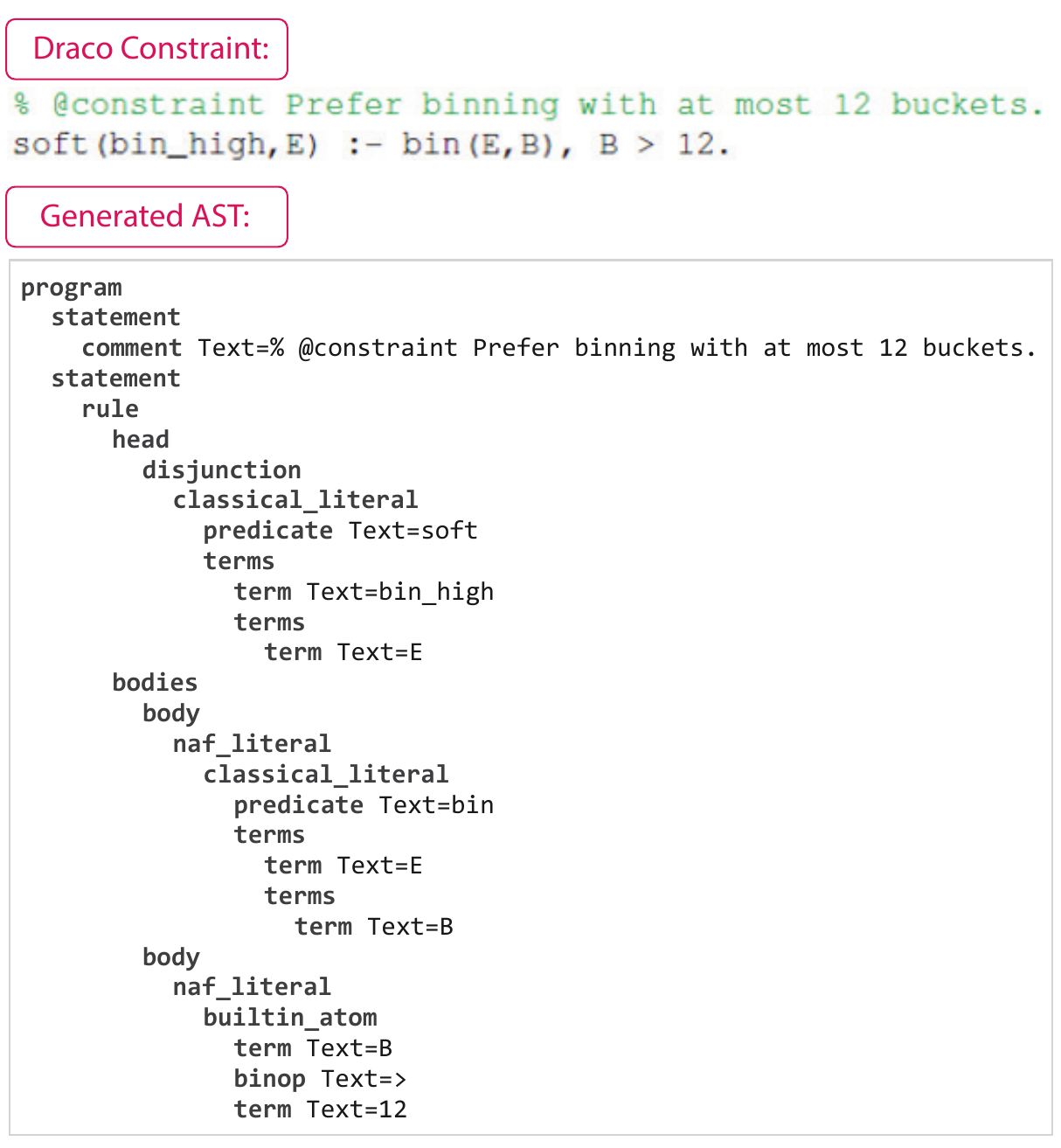}
	\caption{AST generation. A soft constraint with the identifier \emph{bin\_high} is shown, which has a negative impact on the recommended visualization if one of its encodings contains more than $12$ bins. Below the generated AST is shown as a hierarchically decomposed structure.}
	\label{fig:ast}
\end{figure}

\begin{figure*}
	\centering
	\includegraphics[width=\textwidth]{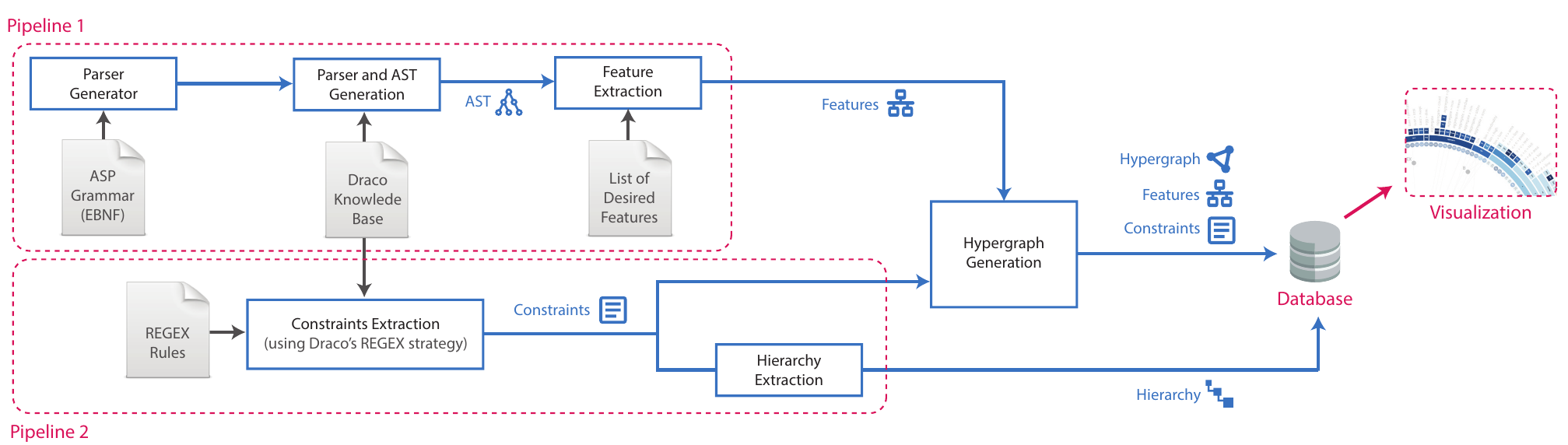}
	\caption{Data extraction. Pipeline 1 extracts the features of Draco's knowledge base and pipeline 2 extracts the constraints, their weights, and the constraints' identifier hierarchy. A hypergraph data generation module processes the features and the constraints and creates the necessary graph-based structure for visualizing a hypergraph of the constraints' shared features.}
	\label{fig:dataextraction}
\end{figure*}

\section{Terms and Definitions}
\label{sec:terms}

In the following we outline the terms and definitions which will be used in the further course of this manuscript. The term definitions partially span from Draco's definition of the knowledge base and our data extraction pipeline, which is explained in \autoref{sec:approach}.

\begin{itemize}
    
    \item \textbf{Facts:} Draco embeds \emph{Vega-Lite}~\cite{7539624} as a formal language to describe its known visualizations. Vega-Lite defines visualizations as a set of mappings from data attributes to graphical elements in a visualization (e.g., x-axis, y-axis, size, color). Draco encodes Vega-Lite visualization building blocks as \emph{logical facts}. Facts describe the properties of the input data and the specified visual encodings (e.g., x-axis mapping).

    \item \textbf{Constraints:} To select proper visualizations (e.g., a bar chart with vertical bars), Draco uses design guidelines, realized as \emph{constraints} over logical facts. The constraints restrict which facts can appear together and which combinations might be preferred. \emph{Hard constraints} prune visual encodings which would result in non-expressive or ill-formed visualization specifications. \emph{Soft constraints} may be violated by a recommendation and determine the final ranking of the recommendation.

    \item \textbf{Knowledge base:} Draco currently holds an (extendable) list of about $230$ hard and soft constraints. Facts and hard and soft constraints together are defined as Draco's \emph{knowledge base}.

    \item \textbf{Features:} We parsed Draco's knowledge base and extracted common elements shared by several constraints as \emph{features}. We later use these features in our VA approach.
    
\end{itemize}

For recommending visualizations, Draco combines the facts and hard and soft constraints stored in the knowledge base with learned weights from a \emph{Support Vector Machine (RankSVM)} model. The model was trained beforehand on labeled visualization pairs. During a recommendation process, Draco processes an input (consisting of a dataset definition, its partial specification, and a user task) to a query definition describing a data schema and query constraints. This query definition is combined with the predefined search space definition consisting of aggregate rules, wellformedness constraints, expressiveness constraints, and the preference model. Draco calls an ASP solver, in this case Clingo~\cite{synthesis/2012Gebser}, to solve the resulting program and subsequently, to obtain ranked answer sets. These answer sets are eventually translated to Vega-Lite specifications. A provided web-based Application Programming Interface (API) of Draco facilitates easy access to the module and decouples Draco from a tool using the system.

\section{Visual Analytics for Rule-Based Systems}
\label{sec:approach}

Before creating a visualization, we had to parse the knowledge base (\autoref{sec:approach:parsing}). Afterward we define requirements for our VA application and explore different design options (\autoref{sec:approach:design}). We propose an interactive, web-based approach (\autoref{sec:approach:framework}) and analyze its usage with respect to specific use cases (\autoref{sec:approach:usecases}).

\subsection{Knowledge Base Parsing and Feature Extraction}
\label{sec:approach:parsing}

Draco's knowledge base is expressed in \emph{Answer Set Programming (ASP)}, a declarative constraint-based logic programming language~\cite{10.5555/3379029}. To organize the knowledge base and make it more readable, Draco's logical expressions have several peculiarities that are not mandatory requested by ASP. Expressions sharing the same goal are grouped in sections. Rules called hard and soft constraints in Draco share the predicates $hard$ and $soft$ in the head of the rule (see also \autoref{fig:ast}, top). The first argument of each soft and hard predicate represents the identifier of the constraint ($bin\_height$ in \autoref{fig:ast}, top). Variables are ASP elements to describe unknown (at the time of definition) individuals. In \autoref{fig:ast} (top), a variable $B$ is used to define the current number of bins. Variables are shortened to a maximum of three letters in Draco. This notation was chosen to increase the readability of the constraints. In total, in the knowledge base, only single-line comments are used.

Although these additional notations are well-considered, ambiguities arise for users who were not involved in the development phase. At first glance, it is unclear what predicates and variables are used in the knowledge base and in which context they are used. Even though the knowledge base developers did their best to find appropriate and expressive names for the predicates and variables, it is not always clear what these arguments are intended for. Their purpose can only be derived by looking at the different contexts where this argument is used. Comments describe the purpose and also the origin of constraints. However, these only constitute textual information. Overlaps of constraints can only be derived through pairwise comparing the constraints.

Hence, we derived that neither the data format of the logical expressions nor their representation is optimal for comprehending Draco's knowledge base and maintaining it. We decided to extract the essential information from the knowledge base for a better picture. For this, we used the ASP parser of Draco and produced an Abstract Syntax Tree (AST) of the knowledge base. The AST extraction provided by Draco's Clingo parser could not be used due to missing Python command-line support. We developed an ASP solver using a parser generator capable of returning an AST of the knowledge base.

Gringo~\cite{synthesis/2012Gebser} is used to parse Draco's knowledge base and ground over it. However, their suggested Extended Backus–Naur form (EBNF) notation of the grammar follows no standard notation style, which is accepted by popular web-based parser generators. Consequently, since we did not find a suitable grammar notation of ASP integrating the syntax features of Gringo, we decided to adapt the grammar of the ASP-Core 2 standard to our needs. This approach is supported and justified, as many introduced syntax features of Gringo are not required to be able to parse the main part of the knowledge base (facts and constraints). The resulting EBNF-based grammar of ASP having the ASP-Core 2 input syntax standard can subsequently be easily used by a parser generator accepting the World Wide Web Consortium (W3C) standard. We used the Open Source code of an EBNF parser generator to generate an easy processable Abstract Syntax Tree (AST) out of Draco's knowledge base. In the generated AST, every node or leaf constitutes a reference to the corresponding terminal definition in the original grammar. We used the AST to extract information about the defined facts and constraints. An example of such an AST can be seen in \autoref{fig:ast}, where an AST generated from a soft constraint is shown.

\begin{figure}[b!]
	\centering
	\includegraphics[width=\columnwidth]{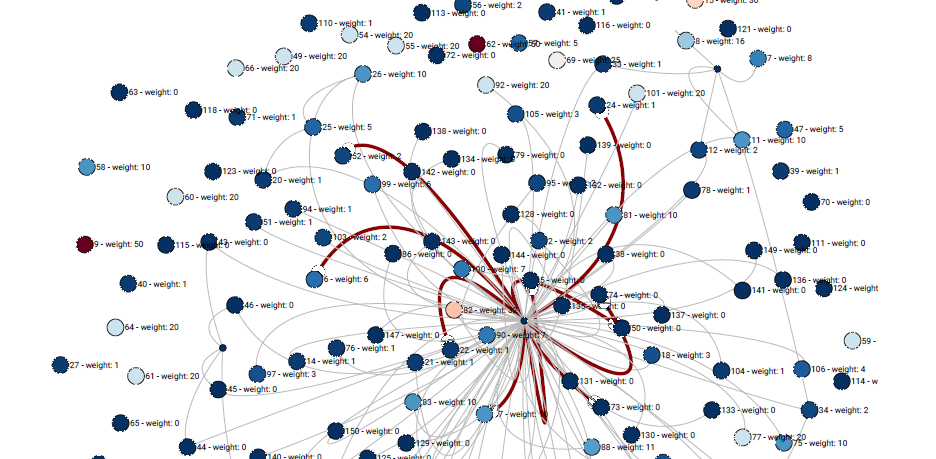}
	\caption{First prototype. Nodes represent constraints and features and edges represent connections between features and constraints. In this first prototype free node-link layout without ordering or attribute-based positioning of the nodes and edges. This led to problems in the analysis, why we decided for a fixed layout later.}
	\label{fig:firstprototype}
\end{figure}

Our full data extraction pipeline is depicted in Figure~\ref{fig:dataextraction}. Generating an AST from the knowledge base enabled us to extract shared syntax elements from the facts and constraints. We refer to these shared elements as \emph{features}. Features were extracted in our data extraction \emph{pipeline $1$}. In a parallel step called data extraction \emph{pipeline $2$}, we pulled the assigned weights for all hard and soft constraints from the knowledge base. We used Draco's own REGEX extraction approach to get this information. Constraints were already annotated in Draco, considerably simplifying this classification and extraction process. The constraints hierarchy could be computed
by processing the constraints’ identifiers resulting from the Draco's own REGEX extraction approach. As a result of both steps, we received shared features (pipeline $1$) and constraints with assigned weights and their hierarchy (pipeline $2$).

In the last step of our parsing and feature extraction pipeline, we united the results from both pipelines into one unique database. We connected constraints based on their shared features. Since features are organized in a hierarchical way, we eventually obtained a hierarchical data structure. We store the entire dataset in a unified database that contains all constraints, features, and the hierarchy of the constraints, using a file-based indexed database.

\subsection{Visualization Design}
\label{sec:approach:design}

The \emph{design triangle} by Aigner and Miksch~\cite{MIKSCH2014286} describes the fundamental aspects influencing the design of VA solutions. It contains the three major key aspects of data, users, and tasks. Based on this definition, we define the following three key elements of our solution:

\begin{itemize}
    
    \item \textbf{Data}: The knowledge base of Draco comprises around $150$ soft constraints and $70$ hard constraints. We extracted shared features among the constraints applying our data extraction pipeline (\autoref{sec:approach:parsing}). The final dataset consists of constraints and shared features stored in one database. 
    
    \item \textbf{Users}: The target users for our VA approach are visualization experts and persons in the research community of automated visualization recommendation and design.
    
    \item \textbf{Tasks}: We identify the following tasks:
    \begin{itemize}
        \item \textbf{T1}: \textit{Analyze the distribution of constraint weights.} This task relates to users being able to better understand the weight distribution within the knowledge base.
        \item \textbf{T2}: \textit{Identify shared syntax features of multiple constraints.} To get an overview of the existing rules, interrelations between the constraints should be made visible.
        \item \textbf{T3}: \textit{Understand violations of recommendations.} For a given recommendation, users should be able to see which constraints have been violated.
        \item \textbf{T4}: \textit{Find recommendations that violate a specific constraint.} For a given constraint, users will need to find out which recommendations violate this constraint.
        \item \textbf{T5}: \textit{Identify recommendations sharing the same violations.} Users want to compare multiple recommendations.
    \end{itemize}
    
\end{itemize}

\begin{figure}[t!]
    \centering
    \includegraphics[width=\columnwidth]{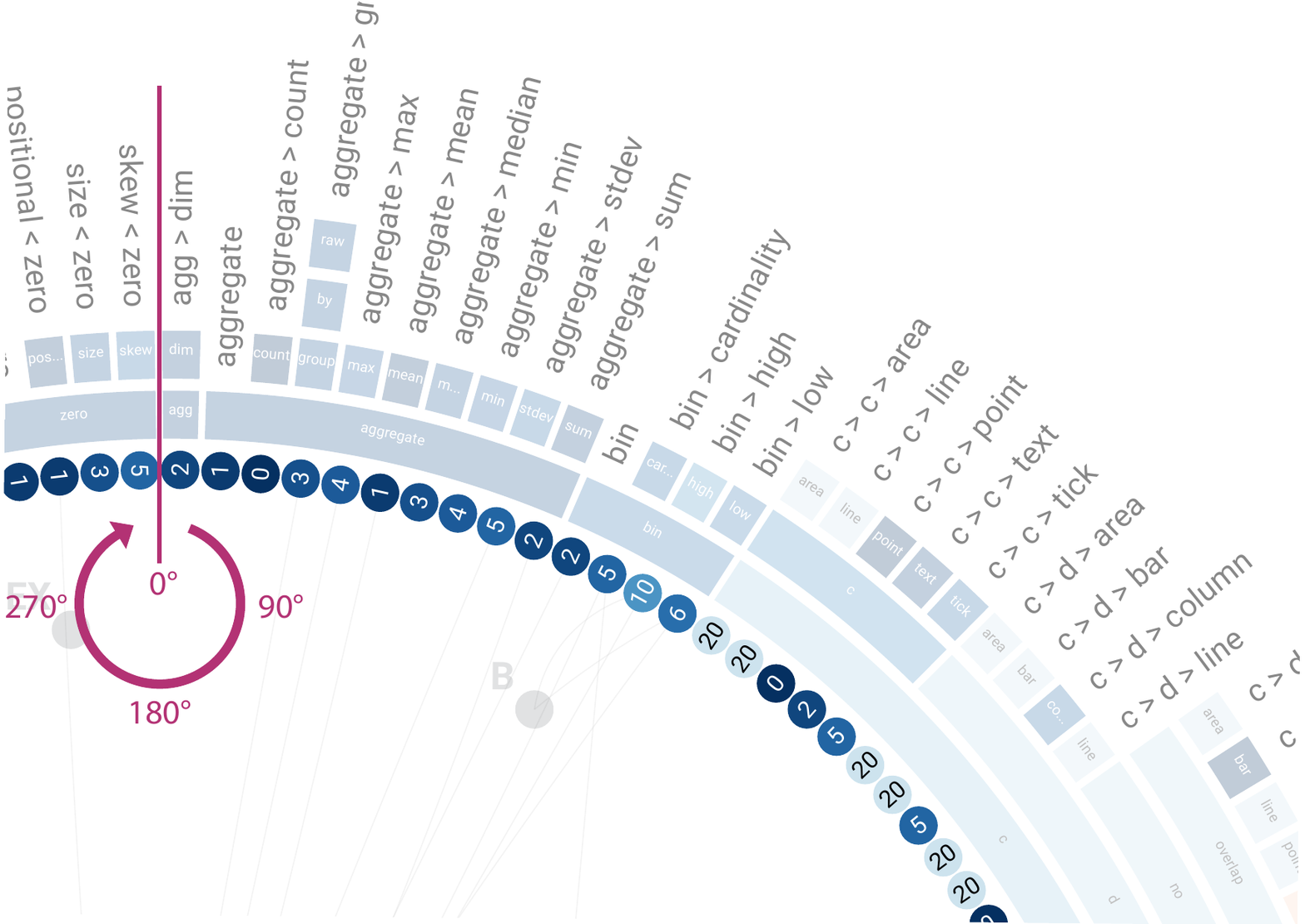}
    \caption{Constraints weights. Constraints are radially arranged in a circle. The constraints' weights are abstracted by round labels which encode the corresponding weight by number and color. Weight labels are rotated according to the angle of the node in relation to the center of the circle.}
    \label{fig:radiallayout}
\end{figure}

The connections between constraints (defined by shared features) can be interpreted as links between constraints. Therefore, as a first approach, we started to apply graph- and network-based visualization techniques. In this prototype, we rendered a network where every constraint represented a node, and every feature-based connection between these constraints (nodes) represented an edge. However, we quickly realized that this kind of edge-mapping results in a very dense graph due to the high number of shared features.

\paragraph{Feature-based connections} To reduce the number of connections within our graph structure, we introduce \textbf{hyperedges}~\cite{6634104}. Visualization designers use hyperedges to reduce the number of edges. For every hyperedge, a new node is created in the graph. In our case, every new node represents a feature. The new nodes are connected to all the constraints sharing this feature. Using hyperedges reduces the number of $e$ edges for $n$ nodes from $O(n*(n-1)/2)$ edges to $O(n)$ edges.

We, again, applied node-link visualization to the resulting \emph{hypergraph}. An initial node-link representation of the hypergraph with a free node-link layout without ordering or attribute-based positioning of the nodes and edges can be seen in \autoref{fig:firstprototype}. Obtaining meaningful structures or orders from the graph was difficult. Due to the flexible layout, each re-drawing cycle of this hypergraph led to repositioning the nodes and edges. It was hard to detect related nodes in this unstructured free node-link layout, and perform any attribute-based sorting of the nodes was infeasible.

\begin{figure}[t!]
	\centering
	\includegraphics[width=\columnwidth]{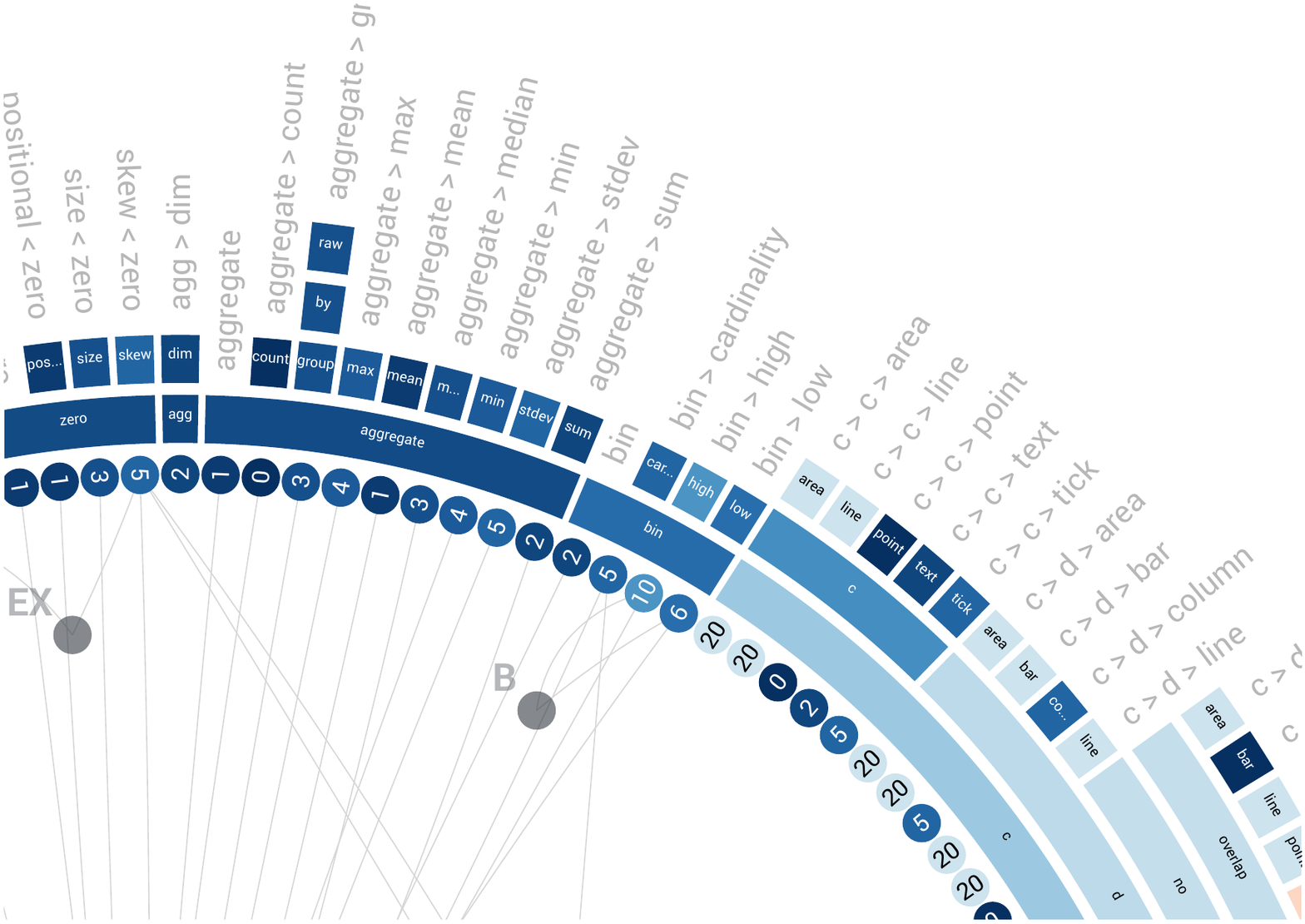}
	\caption{Constraints hierarchy. Circular attached arcs show the hierarchical categories of the constraints. We choose an inside-out strategy where the root hierarchies are placed on the innermost circle. Nodes are colored according to their average weight.}
	\label{fig:hierarchies}
\end{figure}

\paragraph{Node positioning} To fix the positioning of the nodes, we add positioning restrictions rules to our visualization. These rules ensure that the nodes retain their position and the context for every re-drawing cycle of the graph. Nobre et al.~\cite{2019_eurovis_mvnv} distinguished between \emph{styled} layouts and \emph{fixed} layouts. A fixed layout is not applicable, as the constraints do not have any attributes describing an $x/y$ position in a certain space. Hence, a styled layout is required that adds a positioning pattern to the nodes, like arranging them in a line or along a circle. Although the linear arrangement resembles the natural 1D line-based arrangement of the constraints in their code base, many connections between the nodes would lead to a cluttered visualization. We select a \textbf{radial layout}, as it best illustrates a self-contained system where the order of the constraints is not decisive. Furthermore, in a radial positioning, the constraints enclose the nodes representing shared features, which best resembles the visual connection between constraints and feature nodes. The radial approach is additionally supported by other related approaches~\cite{10.2312:eurovisstar.20141170}, where space-efficient representations are required. Our radial layout can be seen in \autoref{fig:teaser} (right) and in \autoref{fig:radiallayout} and \autoref{fig:hierarchies}.

\paragraph{Constraint weights} The constraints are represented as \textbf{nodes} around a large circle. In the case of soft constraints, nodes are colored according to their weight (as seen in \autoref{fig:radiallayout} and \autoref{fig:hierarchies}). We choose a diverging colormap between blue and red with white in the middle. The color blue identifies soft constraints with zero weight, and the red color constraints with the maximum weight related to the total range of the weights $0-50$. Since hard constraints have no weight assigned, all hard constraint nodes share the same color. The nodes' weight labels are rotated according to the angle of the node in relation to the center of the circle (see \autoref{fig:radiallayout}). The problem of poorly readable labels caused by their rotation is tackled by mirroring them between the angles of 90 and 270 degrees (supported by similar layouts~\cite{10.1145/2909132.2926072}).

\begin{figure}[t!]
	\centering
	\includegraphics[width=\columnwidth]{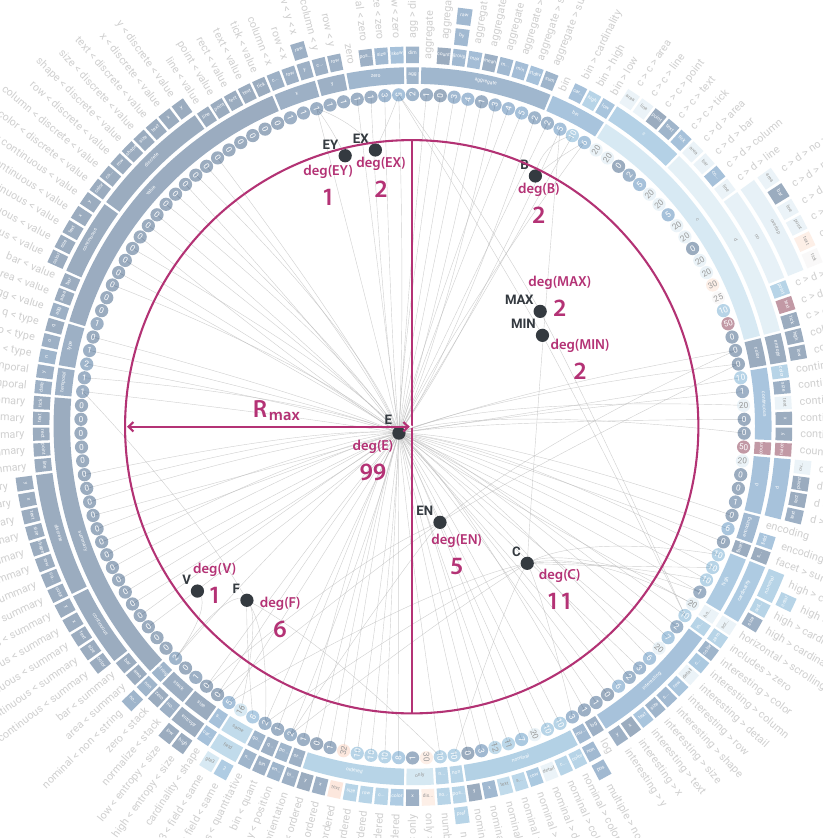}
	\caption{Hypergraph visualization. We place the hypergraph inside the circular arrangement of constraints nodes. Every node in the hypergraph represents a feature and connects constraints which share this feature. Feature nodes are placed according to their centroid calculated from the connected constraints. For example, the position of $EN$ is equal to the centroid $deg(EN)$, which is calculated based on $5$ connections. A maximum radius $R_{max}$ ensures feature nodes to stay inside the circle.}
	\label{fig:hypergraph}
\end{figure}

\paragraph{Constraints hierarchy} Constraints are organized in a hierarchy. There are two possibilities for placing the hierarchical groups next to a graph visualization~\cite{10.1111/cgf.12512}. We distinguish between an \emph{inside-out} and \emph{outside-in} strategy. In the inside-out layout, the root of the hierarchy is placed next to the nodes of the inner circle and the leaf nodes at the outermost circle, depending on the number of levels of the hierarchy. Such a layout is often called a sunburst layout~\cite{8933545}. The outside-in strategy reverses this direction. Leaf nodes of the hierarchy are placed next to the nodes representing the constraints. In Draco, the constraints' identifiers form a hierarchy with variable depths in the leaf levels. Due to this variability, an outside-in layout would introduce empty visual holes toward the center of the graph visualization. We, therefore, employ the \textbf{inside-out strategy} and attach the hierarchical root layers in the innermost circle (see \autoref{fig:hierarchies}). The labels of the constraints are moved outwards so that they are not hidden under the surrounding arcs. The arcs are colored according to the average weight of the respective enclosed constraints. These colored arcs simplify finding groups of constraints with very low or very high weights. They encode their underlying name as a textual label in the arcs' center to identify them. These labels are rotated according to the arcs and shortened when the labels exceed the arcs.

\paragraph{Hypergraph} In our hypergraph, shared features are represented as new nodes connected to the constraints that share these features. Connecting a subset of constraints sharing the same features with simple links results in a fully connected sub-graph. We place the hypergraph \textbf{inside the radial representation} of constraints. Every feature node is visually labeled by the feature's content and connected to every constraint that shares this feature. A feature node's position in the hypergraph depends on its connections to the constraints. We compute a weighted centroid based on the connections to constraint nodes, to which the feature node position converges (\autoref{fig:hypergraph}). Since at least three surrounding constraints always form a regular-shaped polygon, the centroid is invariably inside the outer circle. However, to ensure that feature nodes connecting only one or two nodes fit this positioning restriction, we introduce a maximum radius $R_{max}$ to which feature nodes can converge.

\subsection{Visual Analytics Approach}
\label{sec:approach:framework}

\begin{figure}[b!]
	\centering
	\includegraphics[width=\columnwidth]{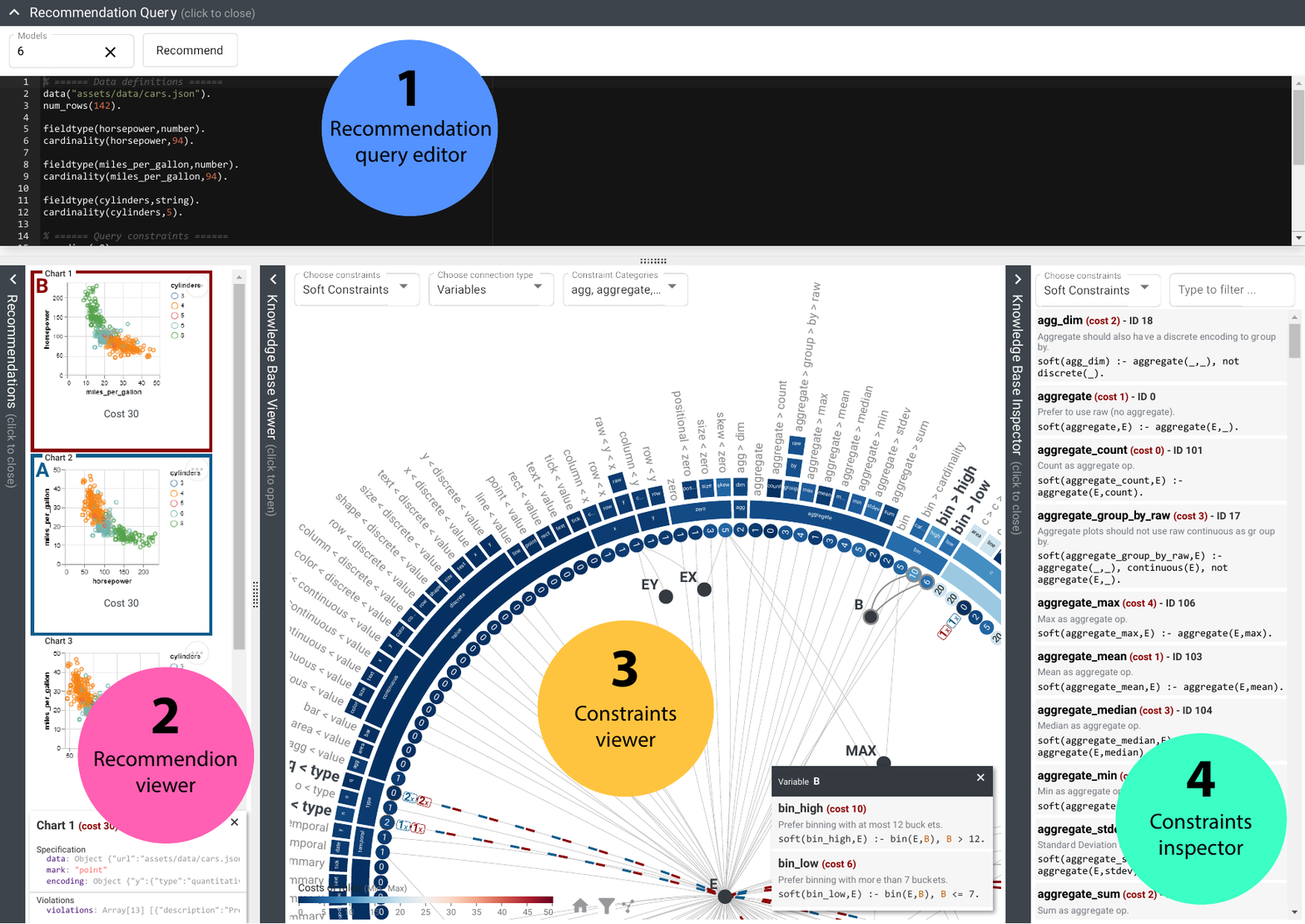}
	\caption{Our proposed VA application. The final application consists of a recommendation query editor (1), a recommendation viewer (2), a constraints viewer (3), and a constraints inspector (4).}
	\label{fig:application}
\end{figure}

We implement our interactive VA approach for exploring Draco's knowledge base based on the described design decisions. The final VA application, which can be seen in \autoref{fig:application}, comprises four major parts:

\begin{enumerate}
	
	\item \textbf{Recommendation query editor}: The recommendation query editor enables users to define an input query to the recommendation engine of Draco. It contains a definition of the data source, its column fields and data types, and statements defining the exploration task regarding the dataset. The query editor allows limiting the number of visualizations to be recommended by Draco.
	
	\item \textbf{Recommendation viewer}: The recommendation viewer shows Draco's recommendations in a tabular view. Depending on the browser window width, multiple columns are shown. The visualizations are sorted row-wise according to their costs (low costs are better). A recommendation's details can be viewed in a tooltip by mouse selection.
	
	\item \textbf{Constraints viewer}: The constraints viewer shows Draco's knowledge base as a hypergraph representation. Multiple controls are available to manipulate the shown data and explore it, for example, switching between hard and soft contraints.
	
	\item \textbf{Constraints inspector}: The constraints inspector shows the raw data of the constraints in a scrollable list view. Two controls allow to switch between the soft and hard controls and to filter respectively search them using a free text search field.
	
\end{enumerate}

Users start an analysis by first asking Draco for recommendations in the \textit{recommendation query editor (1)}. Users can specify a dataset and define which attributes should be visualized here. Afterward, the visualizations recommended by Draco are shown in the \textit{recommendation viewer (2)}. At the same time, Draco's knowledge base is visualized as a hypergraph in the \textit{constraints viewer (3).} Users can now switch between interacting with the recommendations on the left or the hypergraph showing the constraints in the middle. When users are interested in learning more details about specific constraints, these details are shown on mouse click in the \textit{constraints inspector (4).}

The application allows hiding each view independently. Containers are not only collapsable but also resizable. This resizable feature further improves the users' control over the views to optimize the used space regarding the user's task. Following Shneiderman's \emph{Visual Information-Seeking Mantra}~\cite{545307}, users can use the following interactions to filter and explore the data:

\paragraph{Zoom, pan, and filter} The hypergraph visualization (\textit{constraints viewer}) can be freely zoomed and panned by mouse interaction. For zooming, the mouse wheel and/or mouse and keyboard interaction can be used. Panning is done by pressing the left mouse button and mouse movement. Users can revert back to the original visualization state by pressing a button. The hypergraph visualization (\textit{constraints viewer}) incorporates a filter operation by clicking on constraint categories of interest. In addition, above the graph visualization, users can use dropdown boxes to choose between constraint types (\textit{hard} or \textit{soft}) and to select which feature representations should be used (e.g., variables vs. identifiers). We decided to not show hard and soft constraints together, since they define different concepts (i.e., hard constraints cannot be violated, and soft constraints have weights assigned).

\begin{figure}[t!]
	\centering
	\includegraphics[width=\columnwidth]{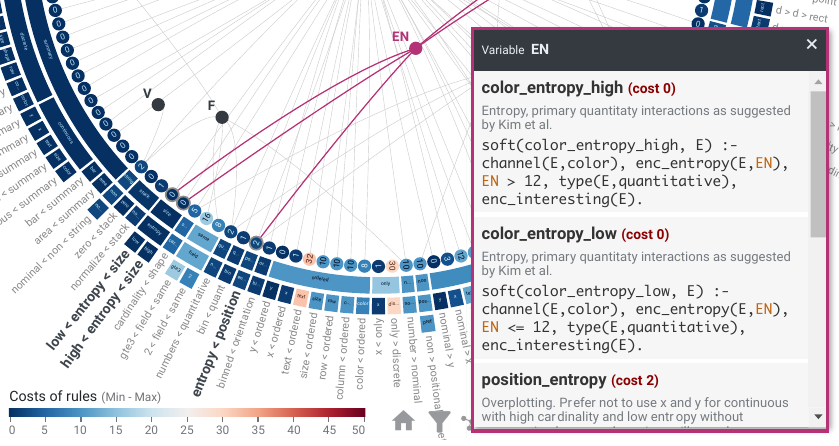}
	\caption{Inspecting feature details. Details on demand for constraints sharing a certain feature of interest that is selected within the visualization. To minimize the overlapped area of the visualization, the tooltip has a fixed position at the bottom right corner of the possible drawing area of the visualization.}
	\label{fig:tooltip}
\end{figure}

\paragraph{Details on demand} Since a user might be interested in the details of a set of constraints sharing a specific feature of interest, a tooltip (\textit{Constraints inspector}) incorporating this information is shown at the bottom right when a feature is selected with mouse interaction (see \autoref{fig:tooltip}). The fixed position at the corner of the visualization reduces the amount of overlap, enabling the enlargement of the tooltip to show more information. The downside is that the visual distance between the clicked feature inside the graph and the tooltip is increased. However, we think that minimized overlaps and showing more information are more beneficial than directly drawing the tooltip next to the selected feature. The selected feature is highlighted.

\paragraph{Brushing and linking} We created links between the \textit{recommendation viewer} and the \textit{constraints viewer}. First, when hovering over a recommendation, the violated constraints are highlighted in the hypergraph. When selecting multiple recommendations, colors are assigned to the selected recommendations which are then reflected by colored stripes in the hypergraph. Whereas framing the recommendations by color works well in the recommendation viewer (colored borders), finding an appropriate highlighting method for the constraints viewer (nodes are tiny) is challenging. Further, one constraint can be violated by multiple recommended visualizations, and one recommendation can violate one constraint several times. We, therefore, propose using a textual label next to the node disclosing the number of violations of each constraint per recommended visualization. Each label shares the same color as its corresponding recommendation. The links to the connected features of the violated constraints are likewise highlighted to determine the affected features by the violations. The problem of overlapping colors is solved using different dash and gap patterns. Our proposed visualization can be seen in \autoref{fig:teaser}. Second, users can hover over constraints in the hypergraph (\textit{constraints viewer}) to see which recommendations violate these constraints.

\subsection{Use-Case Scenarios}
\label{sec:approach:usecases}

Our VA approach was designed with visualization designers as target users in mind (see also \autoref{sec:approach:design}). We defined five tasks that are relevant for this user group: T1 (analyzing the distribution of constraint weights) can be solved by analyzing the displayed weights (numbers and color) in our hypergraph visualization. T2 (identifying shared syntax features of multiple constraints) can be solved by highlighting feature node in the hypergraph. T3 (understanding violations of recommendations) can be solved by hovering over a recommended visualization in the recommendation viewer, which will highlight the respective constraints in the hypergraph. T4 (finding recommendations that violate a specific constraint) can be solved by hovering over constraints in the hypergraph, which will highlight all recommendations violating this constraint. T5 (identifying recommendations sharing the same violations) can be solved by selecting multiple recommendations in the recommendation viewer.

\begin{figure}[t!]
	\centering
	\includegraphics[width=\columnwidth]{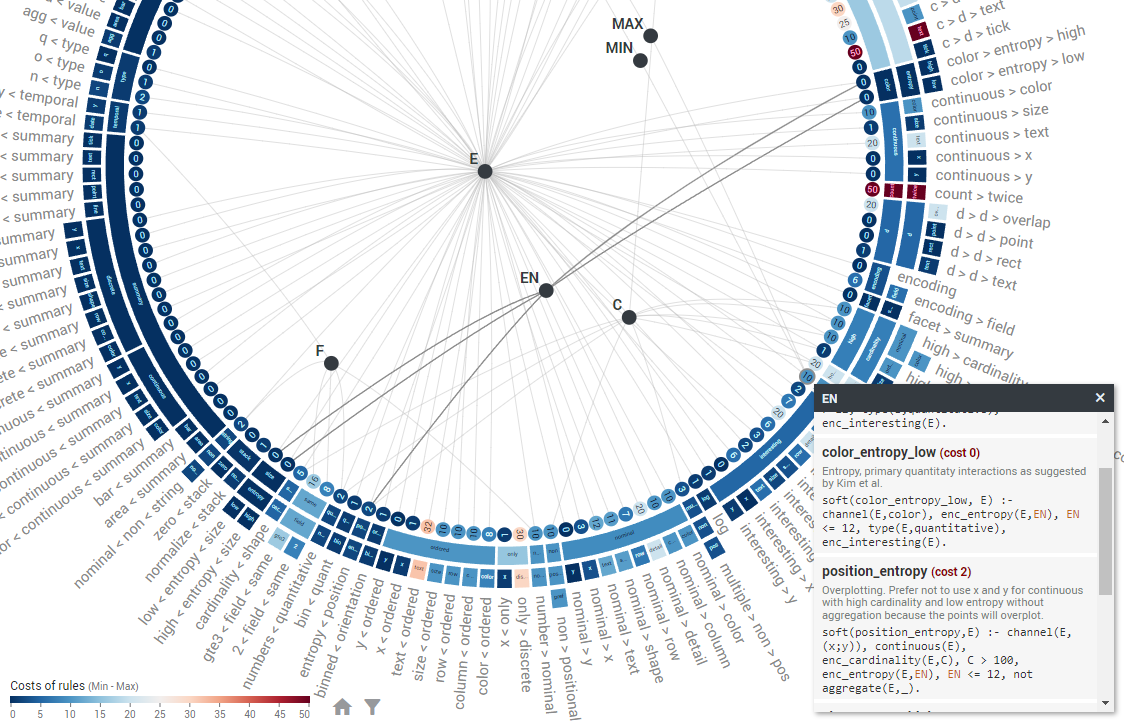}
	\caption{The feature $EN$ refers to entropy and is shared by constraints controlling varieties of colors ($color > entropy$) as well as overplotting ($entropy < position$).}
	\label{fig:usecase1}
\end{figure}

\paragraph{Analyzing the Knowledge Base} The hypergraph (\textit{constraints viewer}) itself provides an overview of the available rules in Draco. Analyzing the interrelationships between constraints is possible by hovering over feature in the hypergraph (\autoref{fig:usecase1}). The feature $E$ (in the background) is shared by almost all constraints in the knowledge base. Other features are only connected to a view constraints. For example, the feature $EN$ (hovered) is shared by constraints controlling variety of color (i.e., not having too many colors in a visualization) as well as constraints describing position entropy (i.e., overplotting issues). As such it is possible to find constraints encoding similar concepts.

\begin{figure}[t!]
	\centering
	\includegraphics[width=\columnwidth]{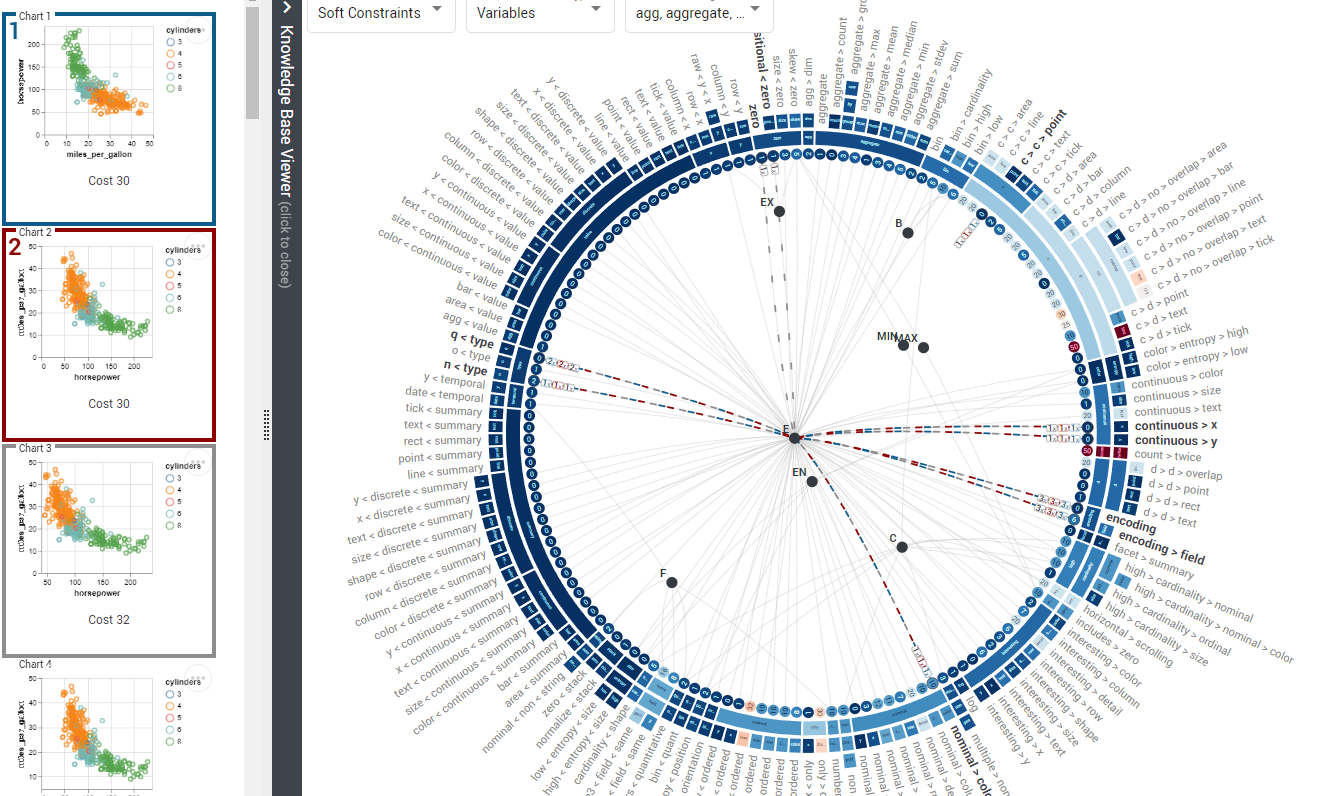}
	\caption{Comparing recommendations. Two recommendations ($A$ and $B$) have been selected, and the VA system assigned them two colors. The violated constraints are highlighted by colored stripes in the hypergraph. When hovering over a third recommendation (grey border), the one constraint ($positional < zero$) not violated by $A$ and $B$ becomes visible, which explains the additional costs of $2$ added to this recommendation.}
	\label{fig:usecase2}
\end{figure}

\paragraph{Analyzing Recommendations} This use case refers to starting with one or a set of recommendations and understanding which constraints have been affected. To compare multiple recommendations, we use different colors to highlight the recommendations and their constraints (T5). Such a use case is shown in \autoref{fig:usecase2}. Here the first ($A$) and the second ($B$) recommendations have been selected by the user. It can be seen that recommendations $A$ and $B$ violate the same constraints. $A$ and $B$ also have the same costs, so they are, in fact, quite similar. When hovering over a third recommendation (grey border), the new recommendation's violations are also highlighted. The third recommendation has higher costs ($32$ instead of $30$) and is linked to a constraint ($poritional < zero$) that is not violated by $A$ and $B$. This shows that the constraint that the x-axis (\textit{horespower} should start with $0$ adds to the costs of the third recommendation.)

\section{Implementation}
\label{sec:implementation}

We rely on a web-based technology stack. Angular is used as the underlying TypeScript-based framework to structure the code base in distinct self-contained parts. The ASP solver Clingo is integrated into the prototype as a compiled WebAssembly (WASM) module. This module is accessed by Draco's web-friendly TypeScript class to execute its knowledge base and, consequently, to generate the recommendations for a given input query. An EBNF parser library is integrated to parse Draco's knowledge base and create the desired AST. The extracted data is persisted in an Indexed Database (IndexedDb) - a widely supported database by modern browsers. For fast prototyping of the desired visualization, the used JavaScript library D3.js is a powerful toolset to manipulate Document Object Model (DOM) elements and to create complex Scalable Vector Graphics (SVG). Angular Material serves as a complementary component and styling library to quickly create common UI components, such as buttons, input fields, lists, and more.
The source code and a demo of our VA approach is available at \url{https://github.com/vrvis/vis_draco_kb}.

\section{Evaluation}
\label{sec:evaluation}

Following the nested evaluation model of Munzner~\cite{5290695}, the evaluation of the visualization's encoding and interaction design is required to assess the usefulness of the proposed approach. To test the visualization regarding its abilities, strengths, weaknesses, and value, we decided to combine four established evaluation methods to get broad feedback:
\begin{itemize}
	\item Observational task performance analysis
	\item Thinking aloud~\cite{lewisusing}
	\item Heuristic-based evaluation methodology (ICE-T)~\cite{8454343}
	\item Qualitative interview
\end{itemize}

These methods are relatively low-cost in their implementation expense and do not require additional equipment (e.g., eye-trackers). Additionally, only a small number of participants are required since studies show that five evaluators are enough to obtain more than $75$ percent of the problems~\cite{8454343}.

The applied evaluation process consists of five consecutive parts. Due to the complexity of the topic and the number of evaluation steps, $45$ minutes were allotted for the entire process:

\begin{enumerate}
	
	\item \textbf{Introduction}: Collecting necessary meta-data of the test person and introducing the user to the topic and evaluation procedure. This first part contained an explanation of the general idea behind visualization recommendation systems, the different existing recommendation systems, the hybrid recommendation system Draco and its knowledge base, and a presentation of the main components of the proposed visualization. Participants had to agree to the evaluation's data recording, privacy, and anonymity terms.
	
	\item \textbf{Free exploration}: The test person could afterward freely explore the visualization and receives answers to upcoming questions. During this exploration, participants were encouraged to ask questions and to think out loud.
	
	\item \textbf{Task performance analysis}: The test person had to solve a set of $12$ pre-defined tasks. Each task started with an introduction to the task, the targeted context, and its goals. As an aid, parts of the tool were hidden that were not required to solve the respective task. While solving a task, the task's description was always visible to the user. The tasks were completed when the test person entered an answer in a designated field and moved on by clicking a button. The tasks comprised the following topics:
	\begin{itemize}
		\item Tasks 1-3: Working with costs (e.g., identifying soft constraints with exceptionally high costs or finding one hierarchy of constraints with very low average weight).
		\item Tasks 4-6: Graph-based questions (e.g., identifying a variable that is part of most soft constraints or naming all constraints using the predicate 'aggregate').
		\item Task 7-10: Understanding recommendations (e.g., identifying how often a given constraint is violated by a recommended visualization or naming at least one constraint by which two recommended visualizations differ).
		\item Task 11-12: Understanding the effect of constraints on recommendations (e.g., identifying how many recommended visualizations are violated by a certain constraint).
	\end{itemize}
	
	\item \textbf{Heuristic evaluation}: After completing the tasks, the test persons were asked to fill out the heuristic value-based survey (ICE-T) by Wall et al.~\cite{8454343}.
	
	\item \textbf{Qualitative interview}: The test persons were asked to give answers to open-ended questions about the tool and the evaluation.
	
\end{enumerate}

The evaluation was conducted via video communication tools. Audio and screen have been recorded. Since web browsers are independent of the operating system and the differences between the available browsers were insignificant, there was no special requirement about it. The test persons were asked to use an external 24-inch monitor with Full High Definition (FHD) resolution. A separate web-based evaluation tool has been created to guide the participants through the evaluation steps and to guarantee a uniform process. Every test person received an evaluation token which allows him/her to enter the evaluation. The token guaranteed that all results could be assigned to the test person while ensuring the person's anonymity.

Since the chosen evaluation methods required only a small number of participants, we decided to pick six test users according to their former knowledge and experience in information visualization. An understanding of basic concepts of  visualization/VA was a requirement in selecting appropriate test persons when considering the degree of specificity of the topic:

\begin{itemize}
	\item P1: 24 years old male master student of Visual Computing having four years of experience in data science and visualization, as well as participated in UX.
	\item P2: 31 years old male who scientifically works in the field of visualization/VA with four years of professional experience. The test person stated that he was already familiar with Draco and its concepts but did not directly work with it.
	\item P3: 39 years old female with a doctoral degree and eight years of experience in data science and visualization/VA.  
	\item P4: 25 years old male master student of Visual Computing having six years of experience in visualization/VA.
	\item P5: 26 years old female master student of data science having three years of experience in visualization/VA.
	\item P6: 32 years old male researcher in the field of biomedical image informatics with 11 years of experience in visualization/VA.
\end{itemize}.

\paragraph{Task performance analysis} Participants did not have problems solving tasks 1-3. Costs and their representation as colors were easy-to-understand and easy to detect elements in the visualization. Similarly, all participants could easily and quickly solve tasks 4-6, which were about interpreting graph connections. For task 7, P2 and P6 were confused by the colored edges connected to the constraint and the number shown in the badge. All participants could solve tasks 8 and 9. For task 10, P1 likely misunderstood the question and declared a constraint on the answer. Tasks 11 and 12 caused the most problems for the participants. P3 and P6 first looked at the problem from the wrong perspective and needed hints to solve the task. P1 had issues understanding the task in the first place but then solved it correctly. P5 also misunderstood the task.

\begin{table}[t!]
	\centering
	\def\arraystretch{1.5}
	\begin{tabular}{r|rrrrrr|c}
		& \textbf{P1} & \textbf{P2} & \textbf{P3} & \textbf{P4} & \textbf{P5} & \textbf{P6} & \textbf{Average} \\ \hline
		\textbf{Insight}    & \cellcolor{Teal!46}4.6 & \cellcolor{Teal!63}6.3 & \cellcolor{Teal!60}6.0 & \cellcolor{Teal!56}5.6 & \cellcolor{Teal!63}6.3 & \cellcolor{Teal!66}6.6 & 5.9 \\
		\textbf{Time}       & \cellcolor{Teal!46}4.6 & \cellcolor{Teal!62}6.2 & \cellcolor{Teal!66}6.6 & \cellcolor{Teal!62}6.2 & \cellcolor{Teal!62}6.2 & \cellcolor{Teal!68}6.8 & 6.1 \\
		\textbf{Essence}    & \cellcolor{Pink!32}3.8 & \cellcolor{Teal!68}6.8 & \cellcolor{Teal!53}5.3 & \cellcolor{Teal!58}5.8 & \cellcolor{Teal!58}5.8 & \cellcolor{Teal!70}7.0 & 5.7 \\
		\textbf{Confidence} & \cellcolor{Teal!43}4.3 & \cellcolor{Teal!67}6.7 & \cellcolor{Teal!60}6.0 & \cellcolor{Pink!37}3.3 & \cellcolor{Pink!37}3.3 & \cellcolor{Teal!47}4.7 & 4.7    
	\end{tabular}
	\caption{Evaluation results. This table shows the evaluation scores according to the evaluation components \emph{Insight}, \emph{Time}, \emph{Essence}, and \emph{Confidence} as defined by Wall et al.~\cite{8454343} (an average value greater as five is considered as success). Participants showed good scores for Time, followed by Insight and Essence. Confidence was rated low, which shows that participants would have to familiarize themselves with the topic and the visualization more.}
	\label{table:evaluation}
\end{table}

\paragraph{Heuristic evaluation} After completing the tasks, the heuristic value-driven survey (ICE-T) as defined by Wall et al.~\cite{8454343} was conducted. In this evaluation scheme, participants rate $21$ statements according to whether they think they are true or false for the proposed visualization. Statements are clustered into the four components of rating \emph{Insight}, \emph{Time}, \emph{Essence}, and \emph{Confidence}. Participants could rate the statements with a grade between $1$ (strongly disagree) and $7$ (strongly agree) or assign $N/A$ if they did not know how to rate the statement. $N/A$ was applied to statement $21$ ('\emph{If there were data issues like unexpected, duplicate, missing, or invalid data, the visualization would highlight those issues.}') by all participants. Apart from statement $21$, participants assigned $N/A$ only in two other cases. The average ratings of the four evaluation components can be seen in \autoref{table:evaluation}. According to Wall et al.~\cite{8454343}, an overall mean of 5 is the threshold of success. The results confirm that our proposed visualization helps users to provide a big picture of the data and an understanding of the data beyond individual data cases. P1 seemed to have problems receiving an overall essence of the data. P4, P5, and P6
worst-rated the confidence component since we received the most feedback from them
regarding applied visual encodings, potential issues, and misleading representations.

\paragraph{Qualitative interview} In the final questionnaire round, all test persons stated that they had a pleasing overall impression of the tool and its visualizations. They mentioned that it is \textit{easy to use, intuitive, and understandable}. P4 pointed out that the visualization especially creates a good overview of which constraints, categories of constraints, and feature-based connections exist within the knowledge base of Draco. P6 noted that the domain and its data are, in general, \textit{very difficult to comprehend} and thus, emphasized the importance of the introduction at the beginning of the evaluation – even for visualization experts. P4 and P5 stated that they had difficulties in using and understanding the filter functionality. They were confused by the checkboxes within the dropdown, and suggested showing all filter possibilities within one view without having to scroll them. P5 also did not expect to be able to select features by clicking on their labels. Sometimes she struggled to select nodes that are placed very close to each other. P4 and P6 suggested unifying the click behavior within the visualization. On the one hand, clicking on the categories leads to filtering the data, and on the other hand, clicking on the features leads to more detailed information. P5 heavily used the constraints inspector during the evaluation; she missed this list's advanced filter and sorting functionality. P2, P4, and P5 stated that the applied colormap from Blue to Red should be changed. The colors used for highlighting the recommended visualizations in the recommendation viewer are the same as applied for constraints’ weights, and they found the diverging colormap misleading. P2 stated that it might convey that constraints having weights in the middle range are neutral.

\paragraph{Summary} Overall, the evaluation has shown that the proposed interactive visualization of the knowledge base viewer is suitable to answer questions regarding the visualization constraints and their shared features. The participants were able to identify the interrelations between the constraints and Draco's recommended visualizations based on highlighted violations. However, the evaluation has also shown that a detailed introduction to the topic is essential, even for participants being familiar with data visualizations. More details about the evaluation can be found in the supplemental material.

\section{Discussion and Future Work}
\label{sec:discussion}

Our VA solution enables to both better understand Draco's knowledge base and its relation to recommended visualizations. We could justify the usefulness of our approach in qualitative and quantitative evaluations. However, both the data extraction approach and the proposed visualization approach still face some limitations:

\begin{itemize}
	\item \emph{Data Processing}: The proposed grammar can only describe selected parts of Draco's knowledge base. For simplicity reasons, Draco's knowledge-based processing methodology is still used to extract the constraints' weights.
	\item \emph{Understanding}: The visualization still requires the user to have a minimal understanding of ASP and the knowledge base behind Draco. Users of the system must be familiar with how Draco encodes visualization facts and how it builds hard and soft constraints over these facts.
	\item \emph{Incompleteness}: The visualization only shows Draco's hard and soft constraints, not the whole knowledge base. The visualization does not depict the visualization facts, optimization statements, and user-dependent input queries.
	\item \emph{Abstraction}: Although the visualization gives an overview of the constraints based on their identifier, categories, and weights, it abstracts their content. As a consequence, the encoded logical expressions can not be inspected directly.
\end{itemize}

The generalizability of the visualization remains to be determined. According to the characteristics of the visualization, it can be used to visualize data structures that consist of a set of entities where the individual entities are part of hierarchical categories. Additionally, the proposed hypergraph within the radially arranged entities can represent the shared features of the entities.

During our research, we identified several possibilities and suggestions for future work in the context of Draco's recommendation system. They concern the proposed constraint viewer in the form of the radial visualization, where it will be interesting to challenge this proposed approach to alternative representations like adjacency matrices, and better representations in node-link diagrams like edge bundling. Our ideas for future work include improving the scalability with ideas for radial sunburst visualization/interaction approaches as proposed by Stasko and Zhang~\cite{885091}. More advanced interaction techniques could be imagined to include more detailed selection options, re-configurations (e.g., swap axis, rotations, re-arranging views), changes on the encoding, and advanced filter operations. Especially eliminating the misleading representations would help to make our prototype more comprehensible. Particularly the confusing highlighting of the violations should be targeted in the future. More research will be needed on how both soft and hard constraints could be visualized together in one view. Future directions might also include using Draco for testing and comparing new theories and empirical study results.

\section{Conclusion}
\label{sec:conclusion}

We presented a VA approach for exploring and understanding a rule-based recommendation system's knowledge base. We specifically focused on visualization recommendation systems, called Draco. Draco has been developed as an automated visualization recommendation system formalizing design knowledge as logical constraints in ASP. With an increasing set of constraints and incorporated design knowledge, even visualization experts lose overview in Draco and struggle to retrace the automated recommendation decisions made by the system. Our VA approach enables visualization experts to relate between Draco's recommendations and constraints stored in the knowledge base and support them in better understanding Draco. We implemented our own data extraction pipeline to extract shared features between constraints from the knowledge base. The resulting shared features and the constraints are then visualized using a hypergraph structure inside a radially-arranged hierarchical visualization. As proven in the evaluation, our approach enables visualization experts to interactively explore the design rules' violations.

\section*{Supplemental Material}

\paragraph{Features and Grammar}
In one document in the supplemental material we added more details about the feature extraction process and outline the structure of the used ASP grammar. This adds to \autoref{sec:approach:parsing} where we explained the data extraction process.

\paragraph{Design Choices}
We added a document to outline representations of our first prototypes. This adds to \autoref{sec:approach:design} where we discussed design choices and decisions of our VA system.

\paragraph{Evaluation}
We added more details and data on the conducted evaluation as described in \autoref{sec:evaluation}. The document contains a more detailed analysis of the results and more information about qualitative feedback.

\acknowledgments{VRVis is funded by BMK, BMAW, Styria, SFG, Tyrol and Vienna Business Agency in the scope of COMET - Competence Centers for Excellent Technologies (879730) which is managed by FFG. Furthermore, this work was funded by the Austrian Science Foundation (FWF) under the grand P31419-N31.}

\bibliographystyle{abbrv-doi-hyperref}
\bibliography{vis2023-final}

\end{document}